\documentclass[jkps,preprint,fleqn,showpacs,showkeys]{revtex4}
\usepackage[pdftex]{graphicx}
\usepackage{amssymb}
\usepackage{amsmath}
\usepackage{bm}
\begin{document}
\setcounter{page}{0}
\title[]{Beam dynamics at the main LEBT of RAON accelerator}
\author{Hyunchang \surname{Jin}}
\email{hcjin@ibs.re.kr}
\author{Ji-Ho \surname{Jang}}
\affiliation{RISP, Institute for Basic Science, Daejeon 135-703, Korea}

\date[]{}

\begin{abstract}
The high-intensity rare-isotope accelerator (RAON) of the Rare Isotope Science Project (RISP) in Daejeon, Korea, has been designed to accelerate multiple-charge-state beams. The ion beams, which are generated by Electron Cyclotron Resonance Ion Source (ECR-IS), will be transported through the main Low Energy Beam Transport (LEBT) system to the Radio Frequency Quadrupole (RFQ). While passing the beams through LEBT, we should keep the transverse beam size and longitudinal emittance small. Furthermore, the matching of required twiss parameter at the RFQ entrance will be performed by using electro-static quadrupoles at the main LEBT matching section which is from the multi-harmonic buncher (MHB) to the entrance of RFQ. We will briefly review the new aspects of main LEBT lattice and the beam matching at the main LEBT matching section will be presented. In addition, the effects of various errors on the beam orbit and the correction of distorted orbit will be discussed.
\end{abstract}



\maketitle

\section{INTRODUCTION}
\begin{figure}
\includegraphics[width=10.0cm]{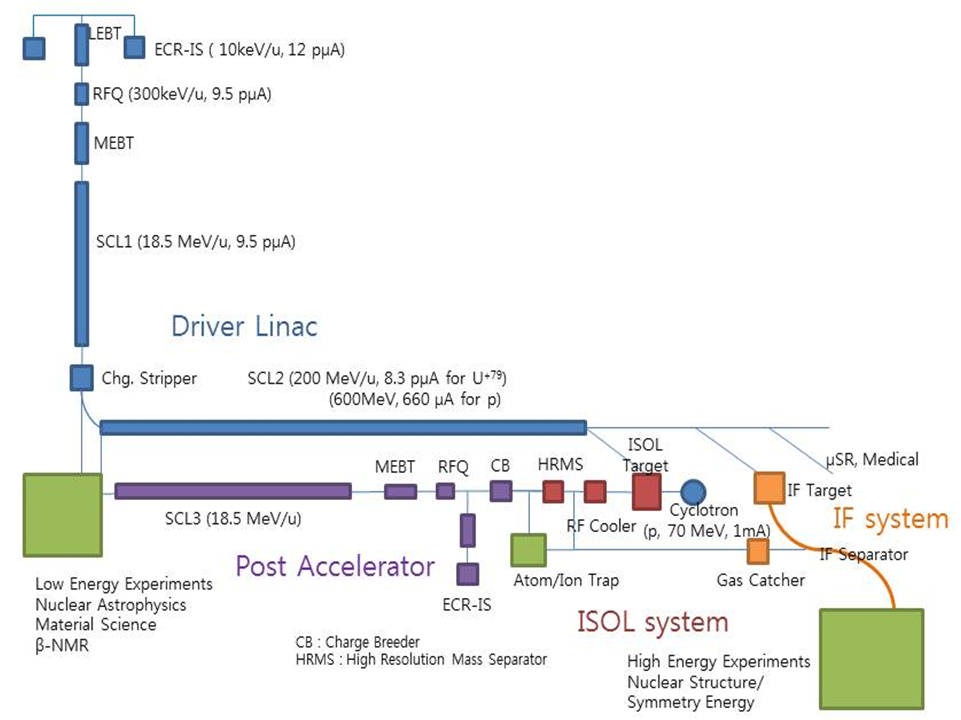}
\caption{Layout of RAON heavy ion accelerator.}
\label{RAON}
\end{figure}

The Rare Isotope Science Project (RISP) was established in December 2011, in the Institute of basic science (IBS) to build a heavy ion accelerator, which is named as RAON (Rare isotope Accelerator Of Newness)~\cite{RAONbase,RAONLINAC,RAONTDR,RAONJKPS} in 2012, for various science programs. The project of RAON heavy ion accelerator is in progress to produce a variety of high-intensity stable ion beams and rare isotope beams to be used for a wide range of basic science researches and various applications in astrophysics, atomic physics, material science, nuclear science and so on.

In the RAON heavy ion accelerator, the in-flight fragmentation (IF) and Isotope Separation On-Line (ISOL)~\cite{ISOL} methods will be used to produce various isotopes. In case of a uranium beam, the final beam energy and power to the various targets will be 200 MeV/u and 400 kW, respectively. To achieve this goal, the IF system uses a diver linac~\cite{SCL1,SCL2} which consists of a 28 GHz superconducting electron cyclotron resonance ion source (ECR-IS), a main low energy beam transport (LEBT) section~\cite{BahngLEBT}, a 300 keV/u RFQ (radio-frequency quadrupole) accelerator, a medium energy beam transport (MEBT) section, a 18.5 MeV/u low energy superconducting linac (SCL1), a charge stripper section (CSS), and a high energy supoerconduction linac (SCL2). Unlike this system, the ISOL system uses a 70 MeV cyclotron as the driver to deliver 70 kW beam power to the ISOL target. The rare isotope beams created by the ISOL system, are reaccelerated by a post linac accelerator which consists of a post LEBT, a post RFQ, a post MEBT, and a 16.5 MeV/u superconducting linac SCL3. The rare isotope beam accelerated by SCL3 will be delivered to the low energy experimental hall or to the SCL2 after passing through P2DT (post-to-driver transport) line. The schematic layout of RAON accelerator is shown in Fig.~\ref{RAON}.

\begin{figure}
\includegraphics[width=10.0cm]{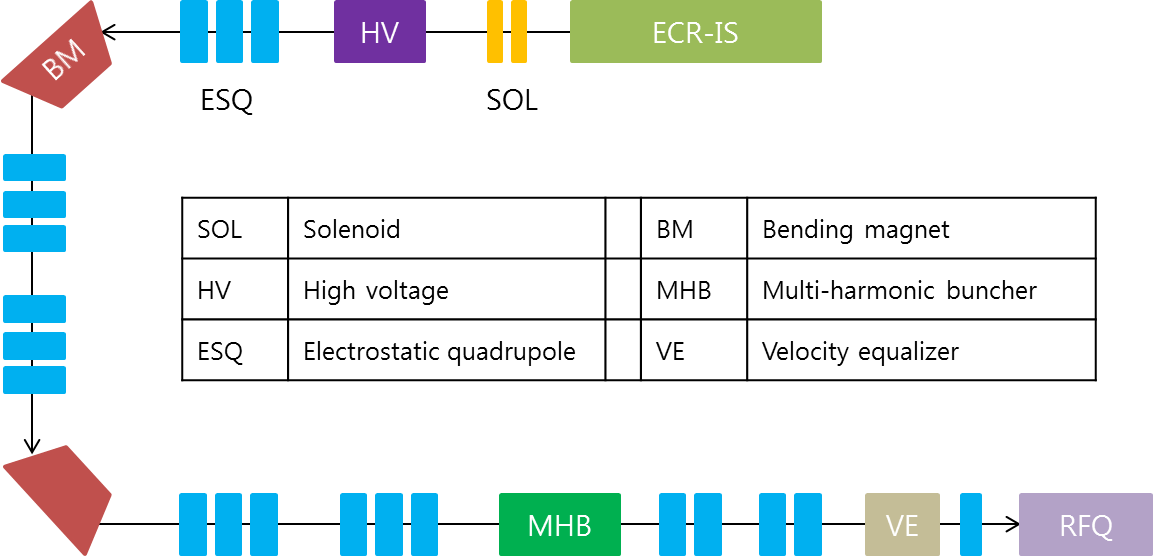}
\caption{Layout of main LEBT.}
\label{LEBT}
\end{figure}

Recently, the lattice of main LEBT is partly modified to decrease the transverse beam size and improve the longitudinal acceptance at the SCL1. The simplified layout is shown in Fig.~\ref{LEBT}. The roles of main LEBT system are to select targeted heavy-ion beams among various ion beams generated by ECR-IS and to deliver these seleted beams to the RFQ with keeping the transverse beam size and longitudinal emittance small. Among various charge-state beams created by ion source, only one or two charge state beams have to be transported to RFQ to match the requirements of beamline experiment. This charge selection is carried out between two 90-degree bending magnets which provide horizontally dispersive region. In the dispersive section, each beam trajectory is changed by the momentum difference of each particle which is directly related to the charge of mass ratio. Selected beams are longitudinally bunched by a multi-harmonic buncher (MHB)~\cite{MHB1,MHB2} to increase the longitudinal transmission at RFQ and SCL1. If two-charge beams are transported, the difference of particle velocities has to be equalized by VE before entering to RFQ. In addition, the twiss parameters have to be matched to the required values at the RFQ entrance to increase beam transmission. Therefore, electrostatic quadrupoles, which are located between MHB and RFQ, are used to match beam twiss parameters to the requirement of RFQ.

In this paper, we will review the upgraded lattice design of main LEBT and discuss the result of beam dynamics simulation with a particle tracking code, TRACK~\cite{TRACK}. At first, the new layout and field map of MHB will be discussed. Next, the matching of twiss parameter at the entrance of RFQ will be described. Finally, the effects of errors induced by magnets throughout the main LEBT will be discussed and the correction of those errors will be presented with limited number of BPMs and correctors.

\section{Beam dynamics in the main LEBT}

\subsection{Design of MHB}

\begin{figure}
\includegraphics[width=10.0cm]{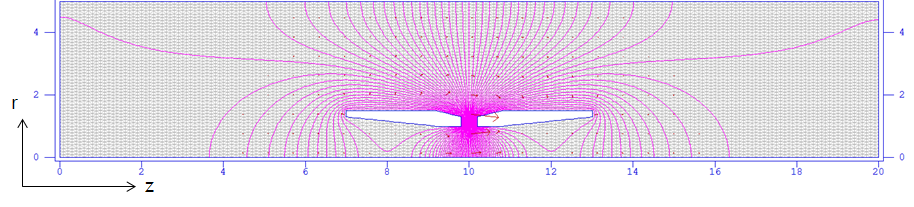}
\caption{Layout of MHB.}
\label{MHB}
\end{figure}

\begin{figure}
\includegraphics[width=10.0cm]{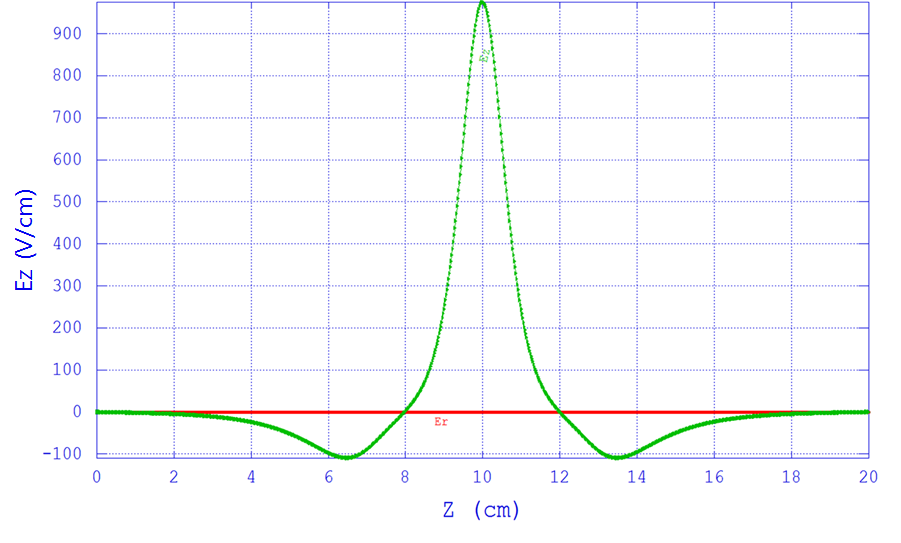}
\caption{Longitudinal axial field map of MHB.}
\label{MHBfield}
\end{figure}

To transport two charge-state beams from ECR-IS to RFQ, beams have to be bunched before entering the RFQ to increase the transmission rate and longitudinal acceptance in the downstream of main LEBT. For that reason, there are two bunchers, a MHB and a velocity equalizer (VE), at the end of main LEBT. Among two bunchers, the MHB is an effective system for bunching low-intensity DC ion beams. The MHB in the main LEBT operates with three frequencies (40.625, 81.25, and 121.875 MHz) to produce an approximately linear sawtooth in the voltage as a function of time to minimize the longitudinal emittance growth. The fundamental frequency of MHB is 40.625 MHz which is half of the RFQ frequency. On the other hand, the VE equalizes the velocity of two beams as a higher energy beam gets lower energy gain than a lower energy beam during passing through VE. The distance between two bunchers is given by~\cite{MHBVE}
\begin{equation}
L_{MHB,VE} = \lambda_{RFQ} \sqrt{\frac{2eV_{0}}{Am_{u}}} \frac{\sqrt{Q_{0}(Q_{0}-1)}}{\sqrt{Q_{0}}-\sqrt{Q_{0}-1}},
\label{eq:MHB}
\end{equation}
where $\lambda_{RFQ}$ is the wavelength of the RFQ, $V_0$ is the accelerating voltage, $A$ is the mass number, $Q_0$ is the hightest charge state of ions.
 
Recently, the design of MHB is modified newly to improve the longitudinal acceptance at SCL1 linear accelerator. Figure~\ref{MHB} shows the side cross-section of the MHB. The total length of MHB is 20 cm and the aperture diameter at middle point is 1 cm. Therefore a maximum transverse beam size should be less than 1cm to avoid beam loss at the MHB. In order to achieve this requirement, three electrostatic quadrupoles ahead of MHB are used. The voltage amplitude of MHB is calculated depending on the particle velocity and distances between MHB, VE, and RFQ. The axial field map of MHB is shown in Fig.~\ref{MHBfield}.

\begin{figure}
\includegraphics[width=10.0cm]{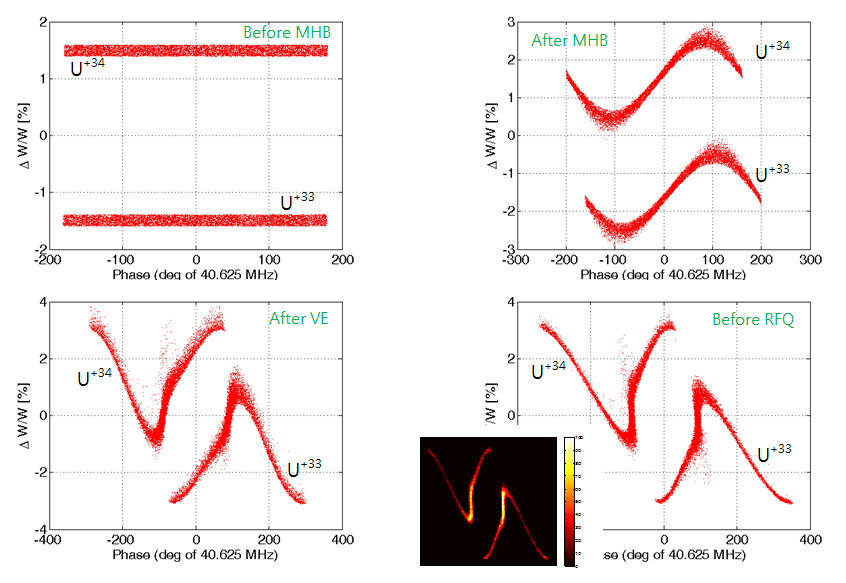}
\caption{Longitudinal particle distributions before MHB (left top), after MHB (right top), after VE (left down), and before RFQ (right down).}
\label{dynamics}
\end{figure}

With newly designed MHB, two charge-state uranium beams, 128U33+ and 128U34+, are tracked with longitudinally uniform distribution as shown in Fig.~\ref{dynamics}. After MHB, beams have saw-tooth energy modulation as keeping energy difference ±1.5 \%. This energy difference disappears after passing through the VE. Consequently, the rms longitudinal emittance is decreased at the SCL1 entrance after modifying the design of MHB. The value of rms longitudinal emittance was 0.057 deg$\times$Mev/u for previous MHB and it decreases to 0.045. Furthermore, this emittance is much smaller than the result of without MHB and VE, 0.096. 

\subsection{Matching twiss parameter}

\begin{figure}
\includegraphics[width=10.0cm]{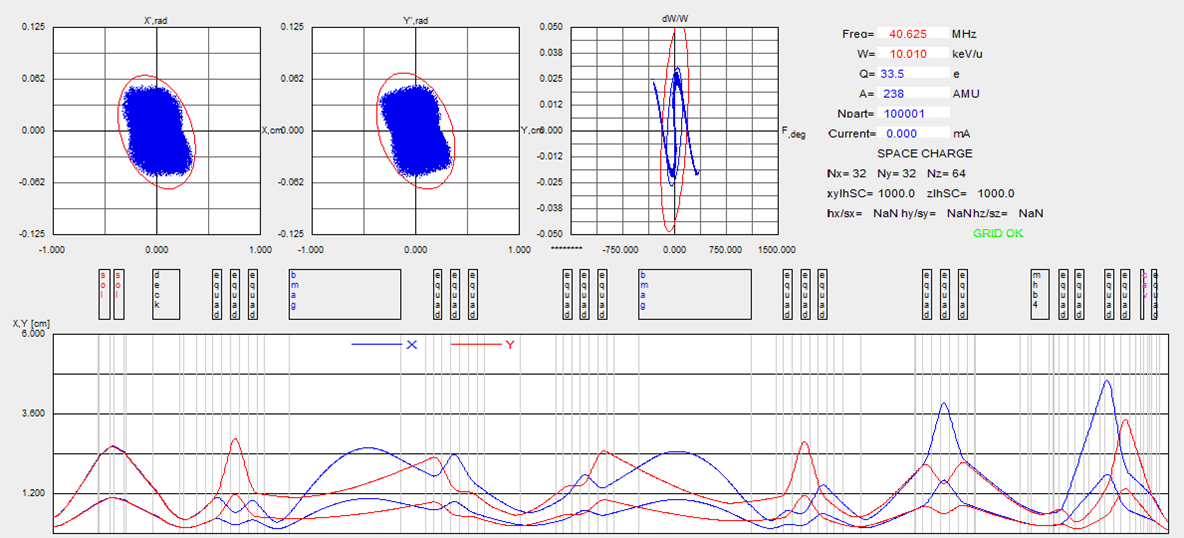}
\caption{Matching of twiss parameters at RFQ entrance with beam envelop in main LEBT. Last five electrostatic quadrupoles are used to match the required twiss parameters at RFQ entrance. Transverse and longitudinal beam profiles at RFQ entrance are shown in the figure.}
\label{matching}
\end{figure}

One of important issues in the main LEBT is the matching of twiss parameters at the RFQ entrance. For higher beam transmission rate, the specific twiss parameters, a betafunction 5.67 cm/rad and an alphafunction 0.324, are required at the RFQ entrance. To do that, at least 4 electrostatic quadrupoles are needed ahead of RFQ. However, a maximum beam size is close to the chamber inner radius after matching with 4 electrostatic quadrupoles. For that reason, 5 electrostatic quadrupoles are used between MHB and RFQ to fulfill all requirements because more quadrupoles can not be installed by reason of limited spaces between MHB and RFQ. Twiss parameters are matched by using simulation codes ELEGANT~\cite{ELEGANT} and TRACK. Finding a matching solution with single particle beam is performed using ELEGANT, and then the matching is carried out a multi-particle beam using a TRACK code. Figure~\ref{matching} shows the TRACK simulation result with a uranium beam, 238U33.5+ after finishing twiss parameter matching with 100,000 macro-particles. As a result, the matching of twiss parameters is performed with high confidence, that is less than 0.1 \% error at the RFQ entrance.

\section{Error analysis}
There are many components which can distort beam orbits in the main LEBT like magnet misalignment, field error, tilt and so forth. Each error component is listed in Table~\ref{error_tol} and the tolerance of those errors is also examined to check the sensitivity of each error source. Tolerance is calculted with a particle loss during simulations of 100 random seeds for each error. Among them, electrostatic quadrupole errors are a main source of beam distortion because there are 20 electrostatic quadrupoles in the main LEBT. 

\begin{table}
\caption{Tolerance of error sources.}
\begin{ruledtabular}
\begin{tabular}{llcc}
 - & - & Value & Unit \\
\colrule
 Initial position & x/y & 0.3/0.5 & cm \\
 Initial angle & xp/yp & 5.0/12.0 & mrad \\
 Solenoid & misalignment & 2.0 & cm \\
 Solenoid & field amplitude & 0.7 & \% \\
 Bending & misalignment & 3.0 & mm \\
 Bending & tilt & 2.5 & mrad \\
 Bending & field amplitude & 0.2 & \% \\
 Quadrupole & misalignment & 0.6 & mm \\
 Quadrupole & tilt & 8.0 & mrad \\
 Quadrupole & field amplitude & 0.15 & \% \\
\end{tabular}
\end{ruledtabular}
\label{error_tol}
\end{table}

\begin{figure}
\includegraphics[width=10.0cm]{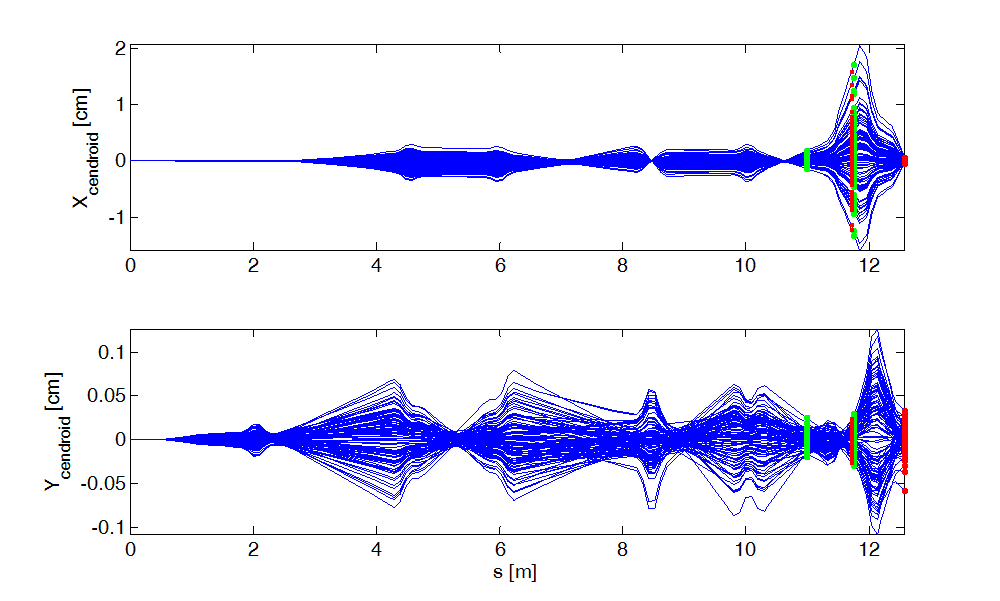}
\caption{Orbit distortion induced by errors listed in Table~\ref{error} with 100 random seeds. Green and red dots mean the position of correctors and BPMs, respectively.}
\label{orbit_error}
\end{figure}

Subsequently, an error study is performed with given errors as listed in Table~\ref{error} for 100 random seeds. The distortion of orbit trajectories is shown in Figure~\ref{orbit_error}. The orbit distortion is most seriously appeared between MHB and RFQ because there are 5 electrostatic quadrupoles with short drift spaces. In the main LEBT, there is no enough space to install lots number of correctors and BPMs, therefore proper number and location of correctors and BPMs have to be investigated. In addition, the distortion of beam orbit should be corrected strictly at the RFQ entrance. Under these limitations, 2 correctors and 2 BPMs are used to correct the distorted orbits. Each corrector is assumed to give a horizontal and vertical kicks at the same time. In Fig.~\ref{orbit_error}, a green and red dots indicate positions of corrector and BPM to be installed, respectively. At a first (second) BPM, a horizontal and vertical rms orbit sizes are 0.5457 (0.0154) and 0.0113 (0.01) cm, respectively. 

\begin{table}
\caption{Errors for orbit correction}
\begin{ruledtabular}
\begin{tabular}{lcccc}
 - & Mialignment [cm] & Tilt [mrad] & Field amplitude [\%] \\
\colrule
 Solenoid & 150 & 5.0 & 0.05 \\
 Bending & 150 & 5.0 & 0.05 \\
 Quadrupole & 150 & 5.0 & 0.05 \\
\end{tabular}
\end{ruledtabular}
\label{error}
\end{table}

\begin{figure}
\includegraphics[width=10.0cm]{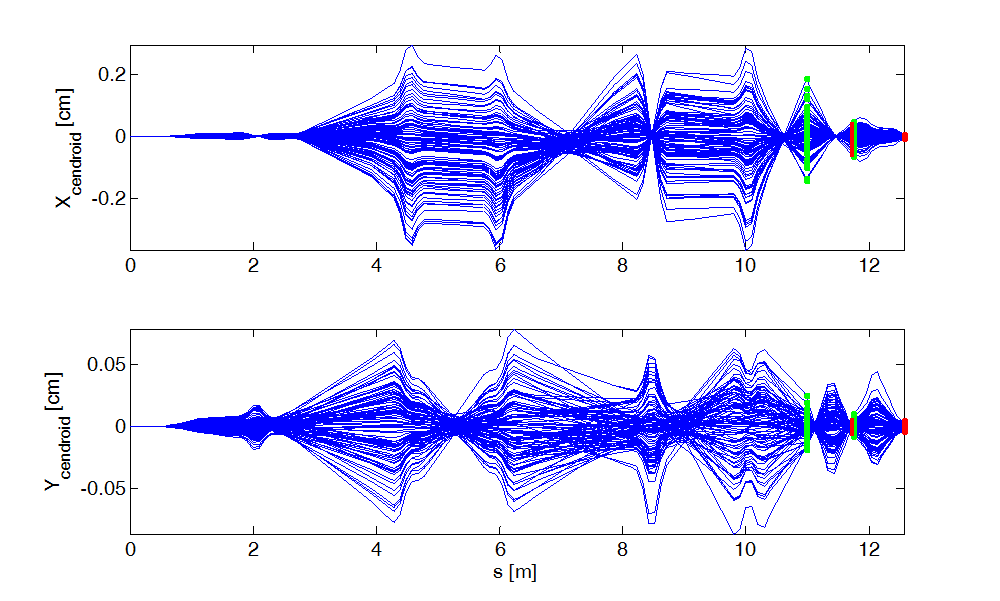}
\caption{Horizontal (top) and vertical (bottom) beam centroids after orbit correction with 2 correctors and 2 BPMs. Green and red dots mean the position of correctors and BPMs, respectively.}
\label{orbit_correction1}
\end{figure}

After correcting distorted orbits, transverse orbit trajectories are shown in Fig~\ref{orbit_correction1}. After correction, the rms orbit sizes at first BPM is 0.0190 cm in horizontal plane and 0.0021 cm in vertical plane, respectively. At second BPM, those are 0.0027 and 0.0019 cm, respectively. The orbit sizes decrease drastically after correction, espically in horizontal plane. During correction of distorted orbits for 100 random seeds, average horizontal and vertical kick angle of first corrector is 2.51 and 0.25 mrad, respectively. The kick angle of second corrector is 0.55 and 0.39 mrad which are less than first one. The calculated kick angle of each corrector is much less than the mechnical limitation, about 7 mrad.

\begin{figure}
\includegraphics[width=10.0cm]{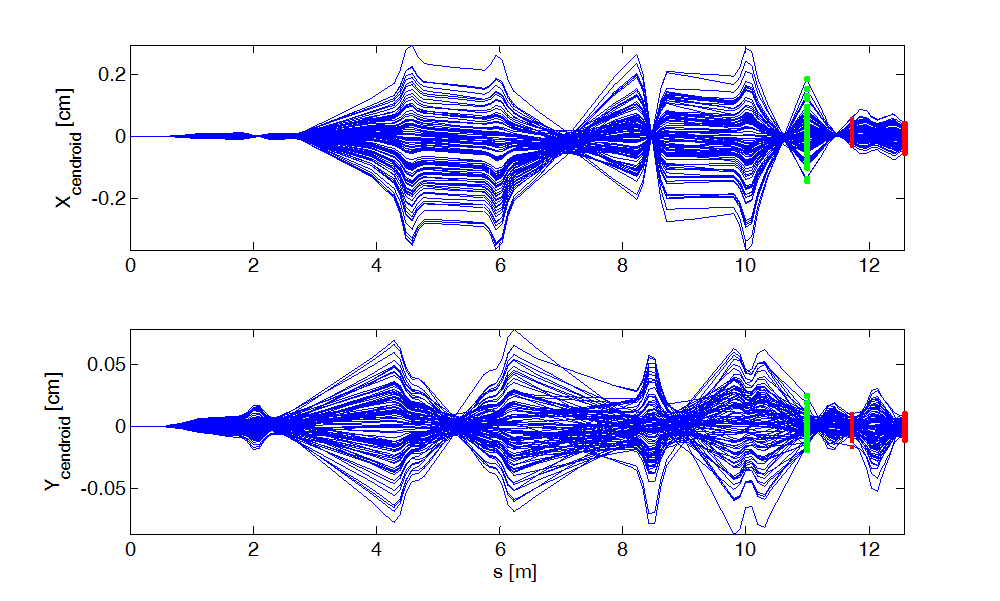}
\caption{Horizontal (top) and vertical (bottom) beam centroids after orbit correction with 1 corrector and 2 BPMs. Green and red dots mean the position of corrector and BPMs, respectively.}
\label{orbit_correction2}
\end{figure}

\begin{table}
\caption{Summary of orbit correction}
\begin{ruledtabular}
\begin{tabular}{lcccc}
 \# of correctors & $<x(y)>_{rms}$ at BPM1 [cm] & $<x(y)>_{rms}$ at BPM2 [cm] \\
\colrule
 0 & 0.5457 (0.0113) & 0.0154 (0.01) \\
 2 & 0.019 (0.0021) & 0.0027 (0.0019) \\
 1 & 0.0187 (0.0047) & 0.0187 (0.0049) \\
\end{tabular}
\end{ruledtabular}
\label{errorsumm}
\end{table}

The orbit correction is carried out with less number of correctors as shown in Fig.~\ref{orbit_correction2}. Without the second corrector, the rms horizontal and vertical orbit sizes at first (second) BPM are 0.0187 (0.0187) and 0.0047 (0.0049), respectively. These values are larger than the above results of 2 correctors, however beam sizes at first BPM are much less than the results of before correction. The result of orbit correction with different number of correctors is summarized in Table~\ref{errorsumm}. Finding proper number and position of correcors and BPMs to correct the effects of errors will be continued.

\section{CONCLUSIONS}
We presented the new aspects of significant beam dynamics issues in the main LEBT of RAON heavy ion accelerator. The design of MHB was modified newly to improve the londitudinal acceptance at the RFQ and SCL1. After changing the design of MHB, the rms longitudinal emittance decreased about 26 \% at the SCL1 entrance. Futhermore, twiss parameters at the RFQ entrance were matched by using 5 electrostatic quadrupoles, accordingly those were matched less than 0.1 \% error with respect to design values. In addition, error studies were carried out in the main LEBT. Significant increase of transverse beam size appeared between MHB and RFQ with magnet errors, and it could be corrected with 2 correctors and 2 BPMs. The rms beam size at BPMs decreased from about 0.5 cm to 0.02 cm after orbit correction within maximum corrector kick angle. The research of error effects and finding proper number of correctors will be continued.

\begin{acknowledgments}
This work was supported by the Rare Isotope Science Project of Institute for Basic Science funded by Ministry of Science, ICT and Future Planning and National Research Foundation of Korea (2013M7A1A1075764).
\end{acknowledgments}

\end{document}